\documentclass[preprint,aps,showpacs,floats,letterpaper,floatfix,groupedaddress,eqsecnum]{revtex4}
\usepackage{amssymb,amsmath}
\usepackage[dvips]{graphicx}

\usepackage{dcolumn,epsfig}

\newcommand{\beq}{\begin{equation}}
\newcommand{\eeq}{\end{equation}}
\newcommand{\bes}{\begin{subequations}}
\newcommand{\ees}{\end{subequations}}
\newcommand{\bea}{\begin{eqnarray}}
\newcommand{\eea}{\end{eqnarray}}
\newcommand{\ba}{\begin{array}}
\newcommand{\ea}{\end{array}}
\newcommand{\beqn}{\begin{eqnarray*}}
\newcommand{\eeqn}{\end{eqnarray*}}

\newcommand{\f}[2]{\frac{#1}{#2}}
\newcommand{\g}{\gamma}

\newcommand{\tc}{\tilde{c}}
\newcommand{\tn}{\tilde{\eta}}
\newcommand{\Se}{\Sigma}
\newcommand{\om}{\omega}

\newcommand{\la}{\langle}
\newcommand{\ra}{\rangle}
\newcommand{\mT}{\mathcal{T}} 
\newcommand{\mF}{\mathcal{F}}

\def\nn{\nonumber}

\newlength{\sizeonefig}
\newlength{\sizetwofig}
\begin{document}

\title{ Electron transport in an open mesoscopic metallic ring}

\author{Dibyendu Roy} 
\email{dibyendu@rri.res.in}
\affiliation{Raman Research Institute, Bangalore 560080}

\begin{abstract}

We  study electron  transport in  a normal-metal  ring modeled  by the
tight binding lattice Hamiltonian, coupled to two electron reservoirs.
First,  B{\"u}ttiker's model  of  incorporating inelastic  scattering,
hence  decoherence and  dissipation, has  been extended  by connecting
each  site of  the  open  ring to  one-dimensional  leads for  uniform
dephasing in  the ring threaded by  magnetic flux.  We  show with this
extension  conductance  remains  symmetric  under flux  reversal,  and
Aharonov-Bohm oscillations with changing  magnetic flux reduce to zero
as a function of  the decoherence parameter, thus indicating dephasing
in  the  ring.  This  extension  enables  us  to find  local  chemical
potential profiles of  the ring sites with changing  magnetic flux and
the decoherence  parameter analogously to the  four probe measurement.
The  local  electrochemical potential  oscillates  in  the ring  sites
because  of quantum-interference effects.   It predicts  that measured
four-point resistance also fluctuates  and even can be negative.  Then
we point out  the role of the closed  ring's electronic eigenstates in
the  persistent current  around Fano  antiresonances of  an asymmetric
open ring for  both ideal leads and tunnel  barriers.  Determining the
real  eigenvalues of  the non-Hermitian  effective Hamiltonian  of the
ring, we show that there  exist discrete bound states in the continuum
of scattering  states for the asymmetric  ring even in  the absence of
magnetic flux.   Our approach involves quantum  Langevin equations and
non-equilibrium Green's functions.

\end{abstract}

\vspace{0.5cm} \date{\today}

\maketitle

\section{Introduction}
 
The  persistent  current in  equilibrium  and  the Aharonov-Bohm  (AB)
oscillations of  conductance with changing magnetic  flux, realised in
normal  metallic ring,  are two  important achievements  of mesoscopic
physics. B{\"u}ttiker,  Imry and  Landauer  \cite{Buttiker83} predicted
the  presence of  persistent  current in  a  closed normal-metal  ring
threaded by a magnetic flux $ \phi $ in the coherent regime.  Magnetic
flux  breaks  down  the  time  reversal  symmetry  of  Schr{\"o}dinger
equation and hence there exists a persistent current whenever the flux
$ \phi  $ is  not equal  to a  multiple of $  {\phi}_0 /  2 $  where $
{\phi}_0  $ is  the  universal  flux quantum.  Gefen,  Imry and  Azbel
\cite{Gefen84} connected  two current leads to  such a one-dimensional
ring and calculated conductance $ G(\phi) $ between the two leads from
the  Landauer formula.   Conductance shows  AB like  oscillations with
changing  magnetic  flux  $\phi$   with  period  $\phi_0$  because  of
inteference  of the  electron  wave-functions coming  through the  two
branches of  the ring  at the  lead.  Another kind  of AB  effect with
principal  period  $\phi_0/2$  is  present  in  the  ring  because  of
interference  of  time  reversed  paths encircling  the  ring.   These
oscillations persist even when strong elastic scattering is present in
the ring.   Both the  persistent current \cite{levy90,chandra91}  in a
closed ring  and the AB oscillations of  conducatance \cite{webb85} of
an  open  ring were  experimentally  realized  at  a few  milli-Kelvin
temperature.

  In  real systems inelastic scatterings are  always present because
of electron-phonon interactions \cite{hod06}  above about 1 K, whereas
electron-electron interactions  are expected to play  dominant role at
low temperatures  in the absence  of extrinsic sources  of decoherence
such   as  magnetic   impurities.    Certainly  inelastic   scattering
introduces decoherence  and both  the above phenomena  are diminished.
B{\"u}ttiker   \cite{Buttiker85,  Buttiker86,   Pilgram06,  Forster07}
proposed  a  phenomenological  model  of inelastic  scattering,  hence
dissipation and dephasing in the  ring. This model is quite similar to
self-consistent reservoirs model, introduced before by Bolsterli, Rich
and  Visscher \cite{Bolsterli70,  AbhiDib06}  in the  context of  heat
transport.   In  B{\"u}ttiker's model,  the  ring  is  connected to  a
reservoir of  electrons of chemical potential  $ \mu $  whose value is
determined  self-consistently by demanding  that the  average electron
current  from the ring  to this  side reservoir  should be  zero. This
conserves the  total number of  electrons in the original  system.  In
this  model the side  reservoir destroys  the coherence  of conducting
electrons  by  removing  them   from  the  transport  channel  and  then
re-injecting them in  the channel with a different  phase and energy; 
thus  dephasing  and  dissipation  can  both  occur.   With  a  single
B{\"u}ttiker probe, conductance of  the open ring enclosing a magnetic
flux satisfies the Onsager reciprocity  relation i.e., $ G( \phi )= G(
-\phi ) $. But in this model dephasing occurs locally in space whereas
in a realistic system it happens uniformly throughout the ring.  There
is  another popular  model \cite{efetov95}  to  incorporate dephasing,
where  a  spatially  uniform  imaginary  potential  is  added  in  the
Hamiltonian of the system which again removes electrons from the phase
coherent  transport channel.  This  model suffers  from a  drawback in
that  it  violates  the  above stated  Onsager  reciprocity  relation.
Brouwer and  Beenakker \cite{brouwer97} have  removed the shortcomings
in the imaginary potential model  by re-inserting back the carriers in
the conducting  channel to conserve  particles. Then they  compare the
two above  stated models for dephasing  in a chaotic  quantum dot.  We
also emphasize that  they consider a single but  many channels voltage
probe.   So  a more  careful  formulation  of  uniform dephasing  with
voltage probes is clearly desirable.

Here we  do a simple extension  to get uniform dephasing  in the ring
with B{\"u}ttiker  probes. All  the sites of  the ring modeled  by the
tight-binding Hamiltonian are  connected with one dimensional electron
reservoirs which  are also  modeled by the  tight-binding Hamiltonian.
Two distant side reservoirs with fixed chemical potentials $\mu_L$ and
$\mu_R$, act as source  and drain respectively. Chemical potentials of
the  other  reservoirs are  fixed  self-consistently  by imposing  the
condition  of zero current.   Now in  this extended  model decoherence
occurs uniformly throughout space.  We show that again the conductance
$ G(\phi)$ is symmetric under flux reversal and the AB oscillations of
$  G(\phi)   $  decay  to  zero   as  the  strength   of  coupling,  $
{\gamma}^{\prime}  $  between the  side  reservoirs  and  the ring  is
increased.  One nice consequence of this extension is that we can find
exact chemical  potantial profiles of  the ring's sites  with changing
magnetic flux by  tuning the coupling $ {\gamma}^{\prime}  $ to almost
zero.  This is similar  to a four-terminal resistance measurement with
non-invasive voltage probes \cite{dePicciotto01}.

Persistent  current in  an  open  ring is  realized  even without  any
magnetic   flux    in   the   presence   of    a   transport   current
\cite{arun95,swarnali04}.   Two  electron  reservoirs  with  different
chemical potentials are  coupled with a mesoscopic ring  in such a way
that the length  of the two arms of the ring  between two contacts are
different. A circulating current flows through the ring around certain
Fermi energy  values where the total  transmission coefficient between
two contacts goes  to zero.  We show here  that at these antiresonance
energy values there exist bound  states in the continuum of scattering
states  (BIC)  for  the  case  of ideal  leads.   For  single  channel
transport bound states'  energies are exactly same as  those of closed
ring's electronic  eigenstates. We also discuss this  issue for tunnel
barriers.

We use the  formalism introduced by Dhar, Shastry  and Sen recently in
two papers  \cite{AbhishekSen06,AbhishekShastry03}.  They have derived
both  the  Landauer results  and  more  generally the  non-equilibrium
Green's function (NEGF) results on transport from the quantum Langevin
equations  approach.   It  is  numerically  easier to  deal  with  the
multiple reservoirs and the disorder in this approach.

The outline  of the paper is  as follows. First we  define the general
model and  describe how  we get different  current expressions  in the
linear   responce   regime   using   quantum  Langevin   approach   in
sec.~(\ref{sec:model}). In sec.~(\ref{sec:ext})  we solve the extended
B{\"u}ttiker's    model    for    uniform    dephasing.     Next    in
sec.~(\ref{sec:Persistent})  we  discuss   the  issues  regarding  the
persistent  current in  an open  asymmetric ring  and bound  states in
continuum.     Finally   we    conclude   with    a    discussion   in
sec.~(\ref{sec:disc}).

\section{Model and general results}
\label{sec:model}

\begin{figure}[t]
\begin{center}
\includegraphics[width=12.0cm]{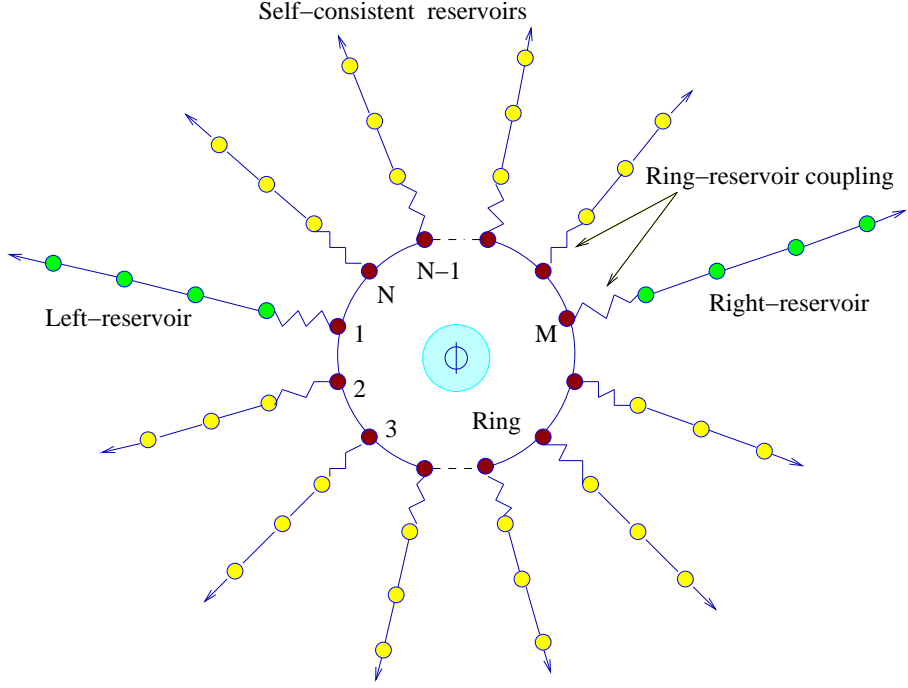}
\end{center}
\caption{ A schematic description of the model. }
\label{cartoon}
\end{figure}

We  consider   a  one-dimensional  mesoscopic  ring   modeled  by  the
tight-binding lattice Hamiltonian. Two distant  sites 1 and $M$ of the
ring are connected to  two infinite reservoirs with specified chemical
potentials $ \mu_1  $ and $ \mu_M $. They  are respectively source and
drain.  Each arm of the open ring between these two contacts has $ N_1
$ and $ N_2 $ sites, each of which is coupled to an infinite reservoir
at  chemical potential  $\mu_l$ and  small finite  temperature $  T $.
[see Fig.~(\ref{cartoon})].  All the  reservoirs are also modeled by a
one-dimensional tight-binding  Hamiltonian.  The total  Hamiltonian of
the system consisting of the ring and all the reservoirs is given by
\begin{eqnarray}
\mathcal{H} &=& \mathcal{H}_r + \sum_{l=1}^N \mathcal{H}_R^l
+\sum_{l=1}^N \mathcal{V}_{rR}^l \nn    \\  
{\rm where}~~~\mathcal{H}_r &=& -\sum_{l=1}^{N}\g~(e^{-i\theta} c^{\dag}_l c_{l+1}+
e^{i \theta} c^{\dag}_{l+1} c_l  )  \nn \\
\mathcal{H}_R^l &=& -\gamma_l~\sum_{\alpha=1}^{\infty}~( c^{l\dag}_{\alpha} 
 c^l_{\alpha+1} +c^{l\dag}_{\alpha +1} c^l_{\alpha})~~~~l=1,2..N \nn \\
\mathcal{V}_{rR}^l &=& - \gamma'_l~ ( c^{l\dag}_{1} c_l + c^{\dag}_l c^l_{1} )~~~~l=1,2...N~. 
\end{eqnarray}
Here $  c_l$ and $ c_\alpha^l  $ denote respectively electron annihilation operators on the
closed ring and on the $l^{th}$ reservoir. Due to periodic geometry of
the ring,  $c_{l}=c_{l+N}$ and contribution  of magnetic flux $  \phi $
has  been included  in  $ \theta=\frac{2\pi  \phi}{N{\phi}_0} $.   The
Hamiltonian of ring is denoted by $\mathcal{H}_r$, that of the $l^{\rm
th}$ reservoir by $\mathcal{H}^l_R$  and the coupling between the ring
and  the $l^{\rm  th}$ reservoir  is $\mathcal{V}^l_{rR}$.   The parameters
$\gamma'_l$  control the  hopping of  electron between  reservoirs and
ring. Also total number of sites in the ring $N = N_1 + N_2 + 2$.

Following Ref.\cite{AbhishekSen06,DibAbhi07}, we  get the steady state
solution of the ring variables in Fourier domain,
\bea
\tc_l(\om)&=&\sum_{m=1}^N ~G^+_{lm}(\om) ~\tn_m(\om) \label{sol} \\
{\rm where}~~~ \tc_l(\om)&=& (1/2\pi) \int_{-\infty}^\infty dt e^{i \om t} c_l(t),~~~G^+= \f{\hbar}{\g}Z^{-1},\nn \\
{\rm and}~~~Z_{lm}&=&\f{\hbar}{\g}(\omega-\Se^+_l) ~\delta_{lm} + e^{- i \theta}\delta_{l,m-1}+ e^{i \theta}\delta_{l,m+1}+ e^{ i \theta}\delta_{l1}\delta_{mN} + e^{- i \theta}\delta_{lN}\delta_{m1}~. \nn  
\eea
 $G^+(\om)$  is the  Green's function  of  the full  system (ring  and
reservoirs) and  for points  on the  ring can be  written in  the form
$G^+(\om)=   [\om-H_r/\hbar-\bar{\Se}^+]^{-1}$   where  $\bar{\Se}^+$,
defined by its matrix elements $\bar{\Se}^+_{lm}=\Se^+_l \delta_{lm}$,
is a  self-energy term modeling  the effect of infinite  reservoirs on
the isolated single particle ring Hamiltonian $H_r$.~ $\Sigma^+_l(t) =
(\f{\g_l'}{\hbar})^2~g^{l+}_{1,1}(t)$  where $g^{l+}_{1,1}(t)$  is the
single particle Green's function of  the lth reservoir at site 1. Here
$\tn(\om)$   is   the   noise   characterising   reservoir's   initial
distribution.  The   effective  ring  Hamiltonian  is   $H_r  +  \hbar
\bar{\Se}^+$ which can  be shown to be non-Hermitian.   We will use it
to find  bound states in a  later section. Now one  important point to
notice is that, for $ \theta $  not equal to an integral multiple of $
\pi  $, $  Z_{lm}  $ is  not  symmetric matrix.   So  the presence  of
magnetic  flux $  \phi $  breaks down  the symmetric  property  of the
$G^+(\om)$ whenever $ \phi $ is not equal to an integral multiple of $
N  {\phi}_0 /2  $.  This  is a  consequence of  the loss  of  the time
reversal symmetry of the problem in the presence of magnetic flux.

In the  present work  we are interested  in electron current  from the
reservoirs to the ring and also current in the ring. For this purpose
we first define  electron density operator on the  ring sites and then
use  the   continuity  equation  to  get   the  corresponding  current
operators. Let us  define $j_l$ as the electron  current between sites
$l~,l+1$ on  the ring and $j_{r-l}$  as the electron  current from the
ring to the  $l^{\rm th}$ reservoir. These are  given by the following
expectation values:
\bea
j_l &=&\f{i e \g}{\hbar} \la ~e^{i\theta} c_{l+1}^\dag c_l- e^{-i\theta}c_l^\dag c_{l+1}~ \ra
\nn \\
j_{r-l} &=& \f{-i e \g_l'}{\hbar} \la~ c_l^\dag c_1^l - {c_1^l}^{ \dag}
c_l ~\ra \nn 
\eea
where $e$ is  the charge of the electron.   Using the general solution
in Eq.~(\ref{sol})  and the  noise-noise correlation \cite{DibAbhi07}, we can do the above averaging 
and find
\bea
j_l&=&\sum_{m=1}^N \f{-1}{2 \pi}\int_{-\infty}^\infty d \om \mF_{lm}~(f_l-f_m)
\label{jer} \\
j_{r-l}&=& \sum_{m=1}^N \f{1}{2 \pi}
\int_{-\infty}^\infty d \om \mT_{lm}~(f_l-f_m) \label{jerl}\\
{\rm with}~~\mF_{lm}&=&\f{2 \pi i e \g {\g'_m}^2}{\hbar^3} ~(~e^{i\theta} G^+_{lm} G^-_{m l+1}-e^{-i\theta} G^+_{l+1 m} G^-_{ml}~)~{\rho}_m \nn \\
{\rm and}~~\mT_{lm}&=&\f{4 \pi^2 e {\g'_l}^2 {\g'_m}^2}{\hbar^4}~|G_{lm}^+|^2 \rho_l \rho_m ~~ ,\nn
\eea
where $G^-_{lm}={G^+_{ml}}^*  $ and $f_l$ is the  Fermi function.  The
chemical potentials  of the reservoirs at  the sites of the  ring 1, M
are specified by  $\mu_1=\mu_L$ and $\mu_M=\mu_R$. Here
we restrict  ourselves at low  temperature and linear  response regime
where      the     applied      chemical      potential     difference
$\Delta\mu=\mu_R-\mu_L$ is small i.e. $ \Delta \mu \ll \mu_{L,R} $ and
$  k_B  T  \ll  \mu_{L,R}$.   For  notational  simplicity  we  choose:
$\g_l=\g$ for  $l=1,2...N$ and $\g'_l=\g'$  for $l=2,3...M-1,M+1,..N$.
With this  assumption, the reservoirs including source  and drain will
have the same  Green's function and density of states  and we will use
the notation $g^{l+}_{1,1}(\om)=g^+(\om)$ and $\rho_l(\om)=\rho(\om)$ \cite{DibAbhi07}.

In the  linear response regime,  taking Taylor expansion of  the Fermi
functions $f(\om,\mu_l,T)$ about the mean value $\mu=(\mu_L+\mu_R)/2$,
Eqs.~(\ref{jer})  and  (\ref{jerl}) reduce  to  the  following set  of
equations:
\bea
j_l &=& \f{-1}{2 \pi \hbar} \sum_{m=1}^N \mF_{lm}~ (\mu_l-\mu_m) \label{jeLR}\\
j_{r-l}&=&\f{1}{2 \pi \hbar} \sum_{m=1}^N \mT_{lm}~(\mu_l-\mu_m)  ~~~{\rm
  for}~~~l=1,2...N~, \label{jerlLR}
\eea 
where  $\mF_{lm}$  and $\mT_{lm}$  are  evaluated at  $\om=\mu/\hbar$.
These are linear equations in $\{ \mu_l \}$ and are straightforward to
solve numerically. In the next section we will consider the case of an
open ring in the presence of uniform dephasing and dissipation. Later,
we will study the persistent current in an asymmetric open ring in the
absence of both magnetic flux and decoherence by external reservoirs.

\section{ Extended B{\"u}ttiker's model for uniform dephasing in  open ring enclosing magnetic flux}
\label{sec:ext}

\begin{figure}[t]
\begin{center}
\includegraphics[width=10.0cm]{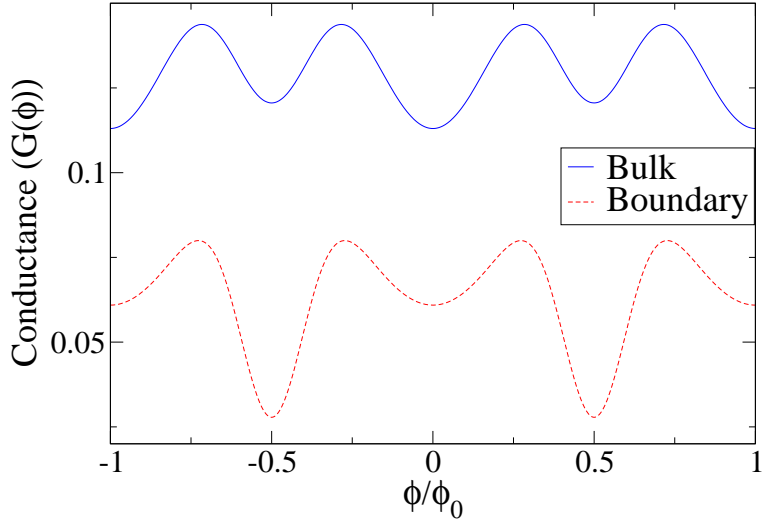}
\end{center}
\caption{Plot of the conductance G($\phi$) of the open symmetric ring with single B{\"u}ttiker probe. The  total number sites in the ring, $N=20$ and ${\gamma}^{\prime}$=1.5 .}
\label{sButt}
\end{figure}

Before  presenting results of  uniform dephasing  in the  open ordered
ring threaded  by magnetic  flux $\phi$, we  first try to  address the
issue of, why  we require an extension of  B{\"u}ttiker's single probe
model,  apart from the  construction of  a more  realistic microscopic
model. In  this section we  work out all  the results for  a symmetric
open ordered  ring, i.e., the number of  sites in the two  arms of the
ring between two contacts at 1  and $M$, are equal, or $N_1=N_2$.  All
the results remain unchanged for the asymmetric case from the physics
point  of view.   Also  we keep  ideal leads  at  1 and  $M$, i.e.,  $
\g_1'=\g_M'=\g$.   We take  a  single B{\"u}ttiker  voltage probe  and
insert it in two  positions of the open ring, once in  the bulk of the
arms between the  two contacts, and then at the  boundary of the arms.
Next the chemical  potential of this voltage probe  is determined from
the self-consistent  condition of  zero average electron  current from
this probe to the ring.  We set from Eq.~(\ref{jerlLR}), $j_{r-l}= 0 $
where  $l$  is the  position  of  the  B{\"u}ttiker probe.   Then  the
equation  is   solved  numerically  for  chemical   potential  of  the
self-consistent  reservoir  with local  density  of  states and  total
Green's function as given in Appendix~\ref{appG}. Finally we calculate
the conductance $  G(\phi) $ between two contacts at 1  and M from the
same Eq.~(\ref{jerlLR}) for  $j_{r-l} $ but with $  l=1~{\rm or}~ M $.
In Fig.~(\ref{sButt})  we plot  $G(\phi)$ with changing  magnetic flux
for two  different postions of the  B{\"u}ttiker probe in  the bulk or
boundary   of  the  open   ring's  arms.    In  both   cases  coupling
${\gamma}^{\prime}$ of  the probe  with the ring  is the  same. Though
conductance  profiles for  the two  above  stated cases  are not  much
different qualitatively  still a  single probe dephases  almost doubly
when  in  the  boundary  than  in  the bulk.   So  there  is  distinct
non-universality  in  the results  from  the  context  of quantity  of
dephasing with a single  B{\"u}ttiker probe depending on it's position
in the ring.

\begin{figure}[t]
\begin{center}
\includegraphics[width=14.0cm]{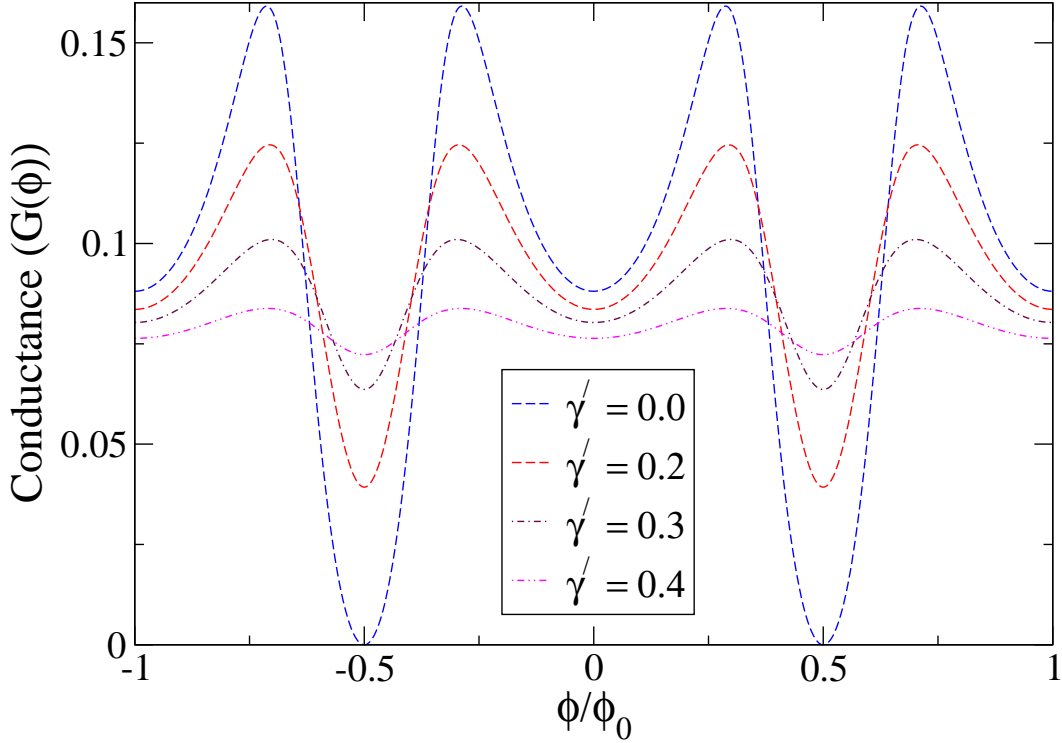}
\end{center}
\caption{Plot of the Aharonov-Bohm oscillations of conductance G($\phi$) of the open symmetric ring with uniform dephasing for different strengths of coupling ${\gamma}^{\prime}$, with $N=20$ .}
\label{ABosc}
\end{figure}

Now we work  out the extended B{\"u}ttiker's model  with all the sites
between  contacts  1 and  $M$  being  coupled  to side  reservoirs  to
simulate other  degrees of freedom present  in a real  ring.  Again to
obtain  the chemical  potentials of  the  side reservoirs  we fix  the
average electron current from these  reservoirs to the ring to be zero
independently. So  we solve the  following $N-2$ linear  equations for
$N-2$ unknown chemical potentials $ \{\mu_l\}$,
\bea
j_{r-l}=0 ~~~{\rm for}~~~l=2,3,...M-1,M+1,...N. \label{SCeq}
\eea 
Once the chemical  potential profile of the side  reservoirs is found,
we  use Eq.~(\ref{jerlLR})  with  $l=1~{\rm or}~M$,  to determine  the
electron current  from the source to  drain. First, we  carry out both
the above jobs numerically. In  all the numerical results presented in
this paper  we set  electrical charge and  Planck constant  $\hbar$ as
unity.   In  Fig.~(\ref{ABosc}) we  plot  conductance  G($\phi$) as  a
function  of  enclosed  magnetic  flux  for different  values  of  the
coupling  ${\gamma}^{\prime}$ of  the side  reservoirs with  the ring.
Here we define conductance as the total current from the source to the
drain   divided  by  chemical   potential  difference   between  them,
$\Delta\mu=\mu_R-\mu_L$.   Clearly   AB  oscillations  of  conductance
G($\phi$)     decay    with    increasing     decoherence    parameter
${\gamma}^{\prime}$   indicating  dephasing.   Also the introduction  of
uniform  dephasing  does not  destroy  Onsager's reciprocity  relation
i.e.,  $ G(  \phi  )= G(  -\phi  ) $.   Using  the similarity  between
different    terms    of     the    full    Green's    function    and
$G_{lm}^+(\om)|_{\phi}=G_{ml}^+(\om)|_{-\phi}$,  we  can  verify  that
under flux reversal the solutions of Eqs.~(\ref{SCeq}) tranform as
\bea
\mu_{l}(\phi)&=&\mu_1+\mu_M-\mu_{l'}(-\phi)~~~~~~{\rm for}~~~~1<l<M,  \\
\mu_{l}(\phi)&=&\mu_1+\mu_M-\mu_{N+l'}(-\phi)~~~~{\rm for}~~~~M<l<N,  
\eea 
where  $l'=M+1-l$.  With   these  transformations  and the above mentioned Green's
function properties, we see that the total current, i.e., conductance, remains
invariant under $\phi \to -\phi$.
\begin{figure}[t]
\begin{center}
\includegraphics[width=14.0cm]{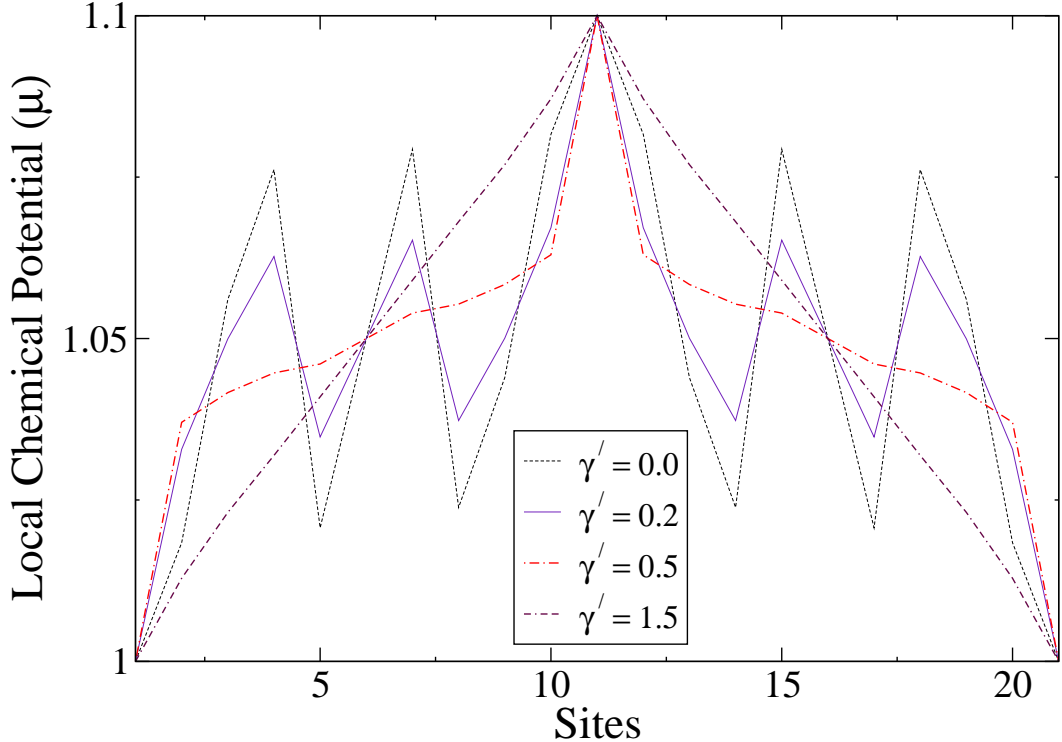}
\end{center}
\caption{Plot of the local chemical potential profiles of the ring sites for different values of the decoherence parameter ${\gamma}^{\prime}$ with ${\phi}$ tends to zero, $N=20$ and site 21$\equiv$ site 1 . }
\label{Chp1}
\end{figure}
\begin{figure}[t]
\begin{center}
\includegraphics[width=14.0cm]{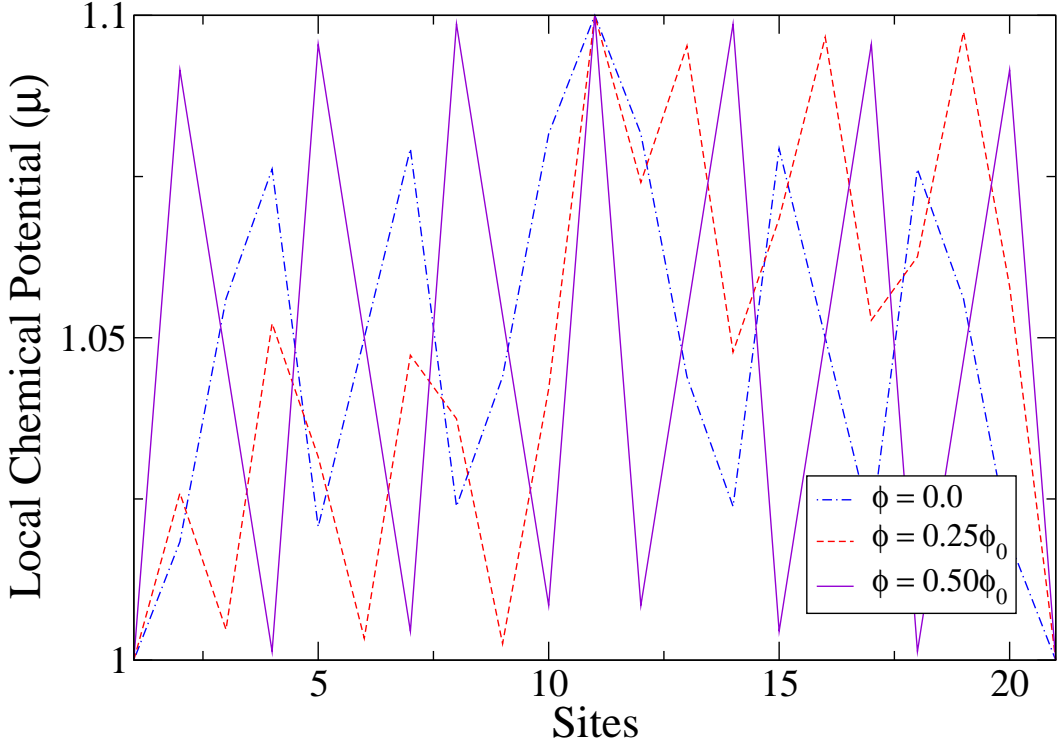}
\end{center}
\caption{Plot of the local chemical potential profiles of the ring sites for different magnetic flux with ${\gamma}^{\prime}$ tends to zero, $N=20$ and site 21$\equiv$ site 1  . }
\label{Chp2}
\end{figure}
As discussed earlier in the  introduction, one elegant outcome of this
extension  is  that, we  can  now  evaluate  local chemical  potential
profiles of  the ring's  sites with changing  magnetic flux  by tuning
${\gamma}^{\prime}$ tends to zero.  This is quite analogous to a four-
probe   measurement  of  a   voltage  drop   in  a   nanoscale  system
\cite{dePicciotto01}.  First we give in Fig.~(\ref{Chp1}) solutions of
the  chemical  potentials from  Eqs.~(\ref{SCeq})  with magnetic  flux
$(\phi)$  tends to  zero.  It  shows large  oscillations in  the local
chemical  potential  profile for  small  ${\gamma}^{\prime}$ and  that
become  more  and   more  flat  with  increasing  ${\gamma}^{\prime}$.
Finally   the   profile    becomes   completely   linear   for   large
${\gamma}^{\prime}$,   signalling   Ohmic   incoherent  transport   of
electrons in this regime, which  has been discussed in great detail in
our  earlier work  \cite{DibAbhi07}.   The oscillations  in the  local
chemical potential profile  for tiny decoherence can be  argued as due
to the  periodic geometry of the   ring. A  electron wave incident
from  the right lead  gives two  contributions to  the current  of the
middle voltage probe measuring local chemical potential. One, there is
direct  transmission into  the probe  and  another, a  portion of  the
carriers  which  are transmitted  past  the  left  lead by  travelling
through the other  arm of the ring and enter the  voltage probe. It is
the  superpostion of these two interfering electron waves which
determines   transmission    in   the   voltage    probe.    Following
M.   B{\"u}ttiker    \cite{Buttiker88,   Buttiker89}   we    call   it
$phase$-$sensistive$ voltage measurement. For slightly larger dephasing,
the flat  behaviour of the chemical  potential profile in  the bulk of
the arms and jumps at the  contacts, is a signature of an intermediate
regime between  ballistic and Ohmic transport.  This  pattern is quite
nicely explained  using a simple  persistent random walk model  in our
previous paper \cite{DibAbhi07}.   In Fig.~(\ref{Chp2}) local chemical
potential profiles of the ring  with changing magnetic flux $\phi$ are
given for the completely coherent case $(\g'=0)$.  For $\phi$ equal to
an integer  multiple of $\phi_0$, the chemical  potential profiles are
the same.   Again for $\phi$ an  integer multiple of  $\phi_0 /2$, the
chemical potential  profiles are similar. In both  cases, the profiles
are symmetric (mirror) about the contacts for the symmetric ring.

Now we derive an analytic expression for the $phase$-$sensitive$ local
chemical potential  profile \cite{Buttiker02}  of the ring  sites with
changing magnetic  flux as in  Fig.~(\ref{Chp2}).  We couple  a single
B{\"u}ttiker probe invasively (though  the final result is insensitive
to the coupling stregth $\g'$) with a middle site of the open ring. We
then determine  the chemical potential  ($\mu_l$) of the  probe, i.e.,
the     corresponding    site,     from     the    self     consistent
Eq.~(\ref{jerlLR}). Moving the probe over all middle sites of the ring
we can evaluate the full $\{\mu_l\}$ profile in a compact form.
\bea
\mu_l=\f{|G_{l1}^+|^2 \mu_L +|G_{lM}^+|^2 \mu_R}{|G_{l1}^+|^2 +|G_{lM}^+|^2}~~~{\rm for}~~~l=2,3,...M-1,M+1,...N,\label{mul}
\eea
where  $G_{l1}^+$  and $G_{lM}^+$  are  given in  Appendix~\ref{appG}.
This derivation will not work for the $\{\mu_l\}$ profile with uniform
finite  decoherence.  The  oscillations  in  the  $\{\mu_l\}$  profile
depends on the Fermi energy  and the applied magnetic flux through the
dispersion relation.

\section{Persistent Current in open asymmetric ring: Role of closed ring's eigenstates   } 
\label{sec:Persistent}

\begin{figure}[t]
\begin{center}
\includegraphics[width=14.0cm]{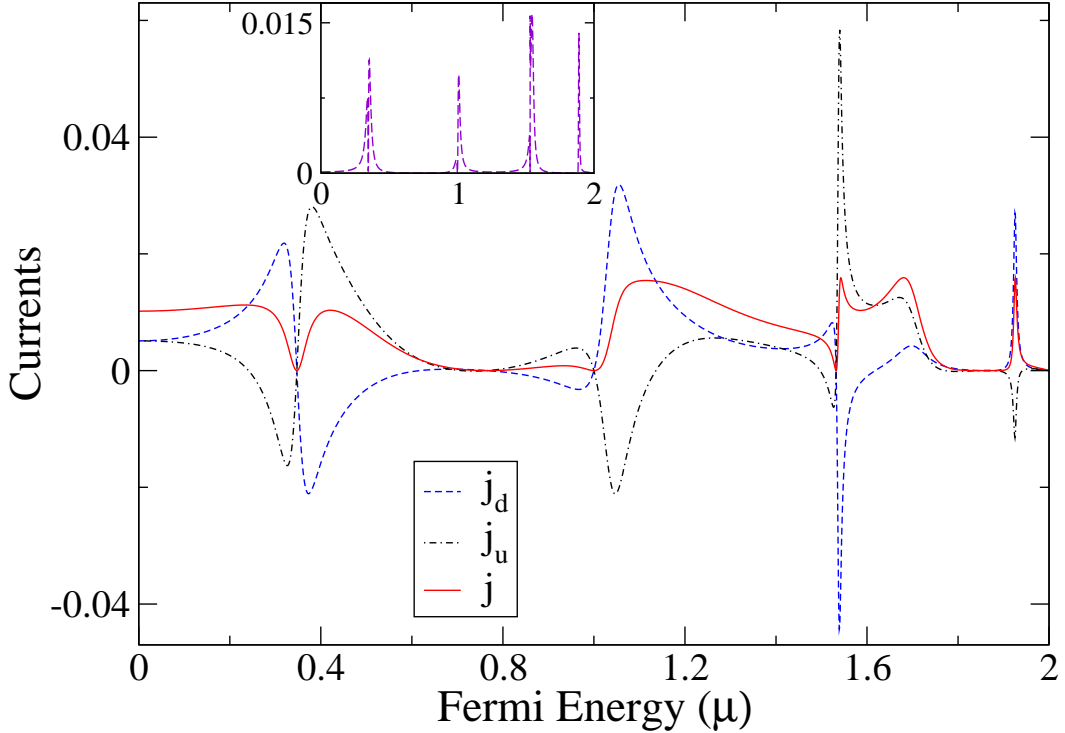}
\end{center}
\caption{Plot of the currents in the two arms $j_u ~,~j_d$ and the total current $j$ in the asymmetric ring with ideal leads $\g''=\g$ in the absence of magnetic flux $\phi$. The inset shows the total current $j$ for tunnel barriers $\g''=0.3 \g$. In both cases $N=18~{\rm and}~N_1:N_2=3:1$~.}
\label{Asymm}
\end{figure}

In  this  section  we  investigate  currents in  a  normal-metal  ring
connected  with  source  and  drain asymmetrically  i.e.   $N_1  \not=
N_2$. Asymmetry is very much required to achieve persistent current or
current  magnification effect  in  the  open ring  in  the absence  of
magnetic  flux.   We  find  an  analytic  expression  for  conductance
$G(\phi)$ between two contacts  of the ring from Eq.~(\ref{jerlLR}) by
evaluating the  Green's function as given  in Appendix~\ref{appG}.  To
find persistent current or circulating current in the ring, we have to
know  currents in  both arms  of the  open ring  separately.  First we
modify  Eq.~(\ref{jeLR})  to  get  current  expressions  $  j_u  ~{\rm
and}~j_d $ in the up and down arms respectively.
\bea
j_u &=& \f{i e \g {\g''}^2}{ \hbar^4}~(~e^{i\theta} G^+_{N M} G^-_{M 1}-e^{-i\theta} G^+_{1 M} G^-_{M N}~){\rho}~(\mu_M-\mu_1) \label{jeU}\\
j_d &=& \f{-i e \g {\g''}^2}{ \hbar^4}~(~e^{i\theta} G^+_{1 M} G^-_{M 2}-e^{-i\theta} G^+_{2 M} G^-_{M 1}~){\rho}~(\mu_M-\mu_1) \label{jeD}
\eea
with $\g'_1=\g'_M=\g''$  and $\g'=0$.  Whenever  current in an  arm of
the asymmetric ring becomes larger  than the total current $j$ between
source  and drain,  there  flows  a circulating  current  in the  ring
exactly equal to  the current in the other arm.   This can be achieved
by tuning  the Fermi energy of  the ring. The phenomenon  of getting a
larger  current in the  arms than  the transport  current is  known as
current magnification.  The  conductance $G(\phi)$ of the normal-metal
ring between two contacts in the presence of magnetic flux is,
\bea
G(\phi)= \f{2\pi e^2 {\rho}^2 {\g''}^4 (-1)^{N-p}{|C^0_{1M}|}^2}{\hbar^5 {|\Delta^0_N|}^2}
\eea
where    $p,~{|C^0_{1M}|}^2~{\rm   and}~|\Delta^0_N|$    are    given   in
Appendix~\ref{appG}.  $G(\phi)$  is defined as $j=G(\phi)\Delta\mu/e$.
Also  we  find  from   the  expressions  in  Appendix~\ref{appG}  that
$G(\phi)=G(-\phi) $.   Similarly exact  expressions of the  currents $
j_u  ~{\rm and}~j_d  $ in  the up  and down  arms of  the ring  can be
evaluated. These expressions are quite long and not included here.  In
Fig.~(\ref{Asymm})  we plot  $  j_u ,~j_d~{\rm  and}~j$ with  changing
Fermi energy of  an asymmetric ring with ideal  leads $\g''=\g$ at the 
two contacts in the absence  of magnetic flux. Clearly for some values
of Fermi energy the current flows through the one arm of the open ring
only whereas  other arm is completely  in an off  condition. There are
some special values of Fermi  energy where the total transmission from
source to drain goes to  zero ($<10^{-10}$). Around these Fermi energy
values  the current  magnification phenomenon  arises.  The asymmetric
pattern of total transmission around these anti-resonances is referred
to as Fano line shape.

We now  determine the energy eigenvalues  of the closed  ring from the
tight-binding  Hamiltonian.   Energy  eigenvalues  are $  E_m  =  -2\g
\cos(2\pi m/N)~{\rm with}~ m=1,2..N$. For $N=18~{\rm and}~\g=1$, there
are  8  doubly  degenerate  eigenvalues,  $\pm$1.87939,  $\pm$1.53209,
$\pm$1, $\pm$0.347296 and  two limiting values $\pm$2 .   We find zero
transmission   points   are  exactly   at   these  doubly   degenerate
eigen-energies.   We suspect  there exist  bound states  of  the total
Hamiltonian  as the transmission  goes to  zero in  the absence  of an
extended  state from  source to  drain. We  will show  below  using an
effective  Hamiltonian  approach that  indeed  they  are bound  states
embedded in  the continuum of scattering states  (BIC). Also different
ratios  between arms'  length, $N_1:N_2$,  do change  the transmission
line shape pattern  but not the antiresonance positions  in the energy
spectrum.  In  the inset of  Fig.~(\ref{Asymm}) we plot  total current
$j$  as  a  function  of  Fermi  energy in  the  weak  coupling  limit
$\g''<\g$.    The  transmission   zeros  at   the   doubly  degenerate
eigenvalues still survive but  the two neighbored resonances around it
almost merge together and their  widths get reduced though the heights
remain  the  same. Also  radiation  shifts  of  the positions  of  the
resonant peaks relative  to the energy eigenvalues of  the closed ring
are  observed in  this regime.   In strong  coupling  limit $\g''>\g$,
whereas  anti-resonance  points  remain  fixed,  the  resonance  peaks
expand.

$Effective~ Hamiltonian~ approach:$ Following Ref. \cite{AbhishekSen06} the bound states are obtained as real solutions of the equation
\bea
[H_r + \hbar {\Se}^+_L(\om) +\hbar {\Se}^+_R(\om)]~|\psi\rangle = \lambda|\psi\rangle,\label{effH}
\eea
where   ${\Se}^+_L(\om)~{\rm   and}~{\Se}^+_R(\om)$  are   self-energy
corrections arising from the interaction of the ring with the left and
right reservoirs respectively. Eigenstates of the tight-binding closed
ring are given as
\bea
|\varphi_m\rangle = \sqrt{\frac{2}{N}}\sum_{j=1}^N \cos(\frac{2\pi mj}{N})|j\rangle.
\eea
Let us multiply $\langle\varphi_m|$ from the left of  both sides of Eq.~(\ref{effH}) and then introduce closure relation $\sum_{n=1}^N|\varphi_n\rangle\langle\varphi_n|=1$ for the isolated ring. 
\bea
\langle\varphi_m|[H_r + \hbar {\Se}^+_L(\om) +\hbar {\Se}^+_R(\om)]\sum_{n=1}^N|\varphi_n\rangle\langle\varphi_n|\psi\rangle = \lambda\langle\varphi_m |\psi\rangle.
\eea
Using the definition of self-energies we get,
\bea
&&\sum_{n=1}^N\langle\varphi_m|\mathcal{H}_{eff}|\varphi_n\rangle \langle\varphi_n |\psi\rangle =\lambda\langle\varphi_m |\psi\rangle~~~{\rm for}~m=1,2...N, \label{seffH}\\
{\rm with}&& \langle\varphi_m|\mathcal{H}_{eff}|\varphi_n\rangle =  E_m \delta_{mn} + g^+(\om)[\frac{{\g'_1}^2}{\hbar}\varphi^{\ast}_m(1)\varphi_n(1)+\frac{{\g'_M}^2}{\hbar}\varphi^{\ast}_m(M)\varphi_n(M)],\nn
\eea
where  $\varphi_n(j)=\langle j |\varphi_n  \rangle$. Eq.~(\ref{seffH})
ia a  matrix eigenvalue equation with  $\mathcal{H}_{eff}$ referred to
as  the non-Hermitian  effective  Hamiltonian in  S-matrix theory  for
transmission \cite{Dittes00,Sardeev03}.  Restricting the energy of the
reservoirs' electron in the  conduction band i.e. $|\hbar\om|<2\g$, we
evaluate the  eigenvalues of Eq.~(\ref{seffH})  numerically.  The real
values of $\lambda$ are precisely the doubly degenerate eigenvalues of
the closed ring.  So there exist bound states at  the energy values of
Fano antiresonances.   Also a change  in the strength or  positions of
the reservoir-ring  couplings does not change the  real eigenvalues of
$\mathcal{H}_{eff}$.  Only the  complex eigenvalues get changed.  This
explains as to why the positions of the antiresonances remain same for
the  different ratios  of the  arms' length,  or with  weak  or strong
couplings.

\section{Discussion}
\label{sec:disc}
In the  present work we have  removed the sensitivity  of dephasing by
the external probe to its position at the bulk and the boundary of the
ring's arm in the B{\"u}ttiker's single probe model, by coupling every
site of  the open ring  with self-consistent reservoirs.  Of  late the
mesoscopic  AB oscillations  have  served as  a  measuring device  for
different mechanisms of electron decoherence such as electron-electron
scattering,     and     scattering     off     magnetic     impurities
\cite{wagner06,Pierre03,Pierre02}.  Our extended  model will be useful
to understand the experiments where the decoherence in the ring occurs
uniformly because of the interactions of conducting electrons with the
other  degrees of  freedom present  in  the system.   There are  other
perfectly valid models for uniform dephasing \cite{arun02, brouwer97}.
Still  our extension  is closer  to experiments  as here  the coupling
between the  ring and  the environment is  direct and  easily tunable.
Recently the  resistance of  single-wall carbon nanotubes  have been
studied  \cite{Gao05} in a  four-probe configuration  with noninvasive
voltage electrodes.   They have  found that the  four-probe resistance
fluctuates and  can even become negative at  cryogenic temperature due
to  quantum-interference  effects   generated  by  elastic  scatterers
\cite{Buttiker89} in the nanotube. With recent progress in experiments
with quantum rings  \cite{Fuhrer01} we believe that it  is possible to
detect the local chemical potential  oscillations in the open ring as
predicted  in   the  present  paper.  Here  we   should  mention  that
differences  between   $phase$-$sensitive$  and  $phase$-$insensitive$
measurements   are   drastic   for   an   effectively   single-channel
transmission  problem  compare  to  multichannel  conductor  where  it
depends    on   the    particular   arrangement    of   probe-coupling
\cite{Buttiker89}.    It  is  also   required  further   attention  to
investigate  effects  of   static  disorder  (elastic  scatterer)  and
electron-electron   interaction  on   the  local   chemical  potential
oscillations.   There is  good scope  to  study the  mutual effect  of
disorder  and  dissipation  in  dissipative open  quantum  systems  by
introducing  disorder in  the  ring Hamiltonian  through our  extended
model in the quantum Langevin equation approach.

The  Fano  antiresonance  occurs  because  of the  interference  of  a
discrete autoionized state  with a continuum.  Here we  have shown for
the  single  channel  transport  in  the  asymmetric  open  ring,  the
antiresonances   occur  exactly  at   the  doubly   degenerate  energy
eigenstates of  the closed  ring in the  absence of  evanescent modes.
Also by  finding the real  eigenvalues of the  non-Hermitian effective
Hamiltonian  we  predict  the  existence  of the  BIC  at  these  Fano
antiresonance energy  values.  Recently  some more studies  have found
the BIC in an AB ring and in a double cavity electron waveguide \cite{
Bulgakov06,  Ordonez06}.  Here  we emphasize  that in  case  of single
channel  transport the total  transmission of  an open  symmetric ring
never goes  to zero  in the absence  of magnetic flux.   Finally bound
states do not contribute directly in the transport for non-interacting
systems.   However  as  suggested  by  a  mean  field  calculation  in
Ref.\cite{AbhishekSen06},  they may affect  the current  via affecting
the local  density in the presence  of electron-electron interactions.
It will  be interesting to see  that how the  interactions between the
electrons affect the transmission zeros in the open asymmetric ring.
 
\section{acknowledgments}

The  author  wants  to thank  Abhishek  Dhar  and  N. Kumar  for  many
discussions on the phenomenon of decoherence, and to R. Srikanth for a
critical  reading of  the manuscript.  The author  would also  like to
express his gratitude to M.   B{\"u}ttiker for his critical remarks on
the  manuscript  and K.   Ensslin  for  a  helpful discussion  on  the
possibility  of  experimental realisation  of  the chemical  potential
oscillations in quantum rings.

\appendix

\section{Evaluation of Green's Function}
\label{appG}

 The      full       Green's      function      is       given      as
$G^+_{lm}=(\hbar/\gamma)Z^{-1}_{lm}$  where $Z$  is  a near  circulant
matrix with off-diagonal  terms $Z_{N 1}= Z_{l~l+1}=e^{-i \theta}~{\rm
for}~l=1,2...N-1$      and     $Z_{1N}=Z_{l-1~l}=e^{i     \theta}~{\rm
for}~l=2,3...N$. The diagonal terms are given by:
\bea
Z_{11}&=&Z^+_{MM}=A(\om)= \f{\hbar}{\g}~[\om-\f{{\g''}^2}{\hbar^2}
  g^+(\om)]~~~{\rm with}~~\g'_1=\g'_M=\g''~, \nn \\
Z_{ll}&=&B(\om)= \f{\hbar}{\g}~[\om-\f{{\g'}^2}{\hbar^2}
  g^+(\om)]  \label{diagelm}~~~{\rm for}~~l=2,3...M-1,M+1...N~.
\eea 
Now using the method of Ref. \cite{DibAbhi07} to determine inverse and determinant of the tri-diagonal matrix, we can find required inverse and determinant of the near circulant matrix $Z$ through simple but tedious algebra. 
\bea
\Delta_N&=&\left((A-2\cosh\alpha)^2(\cosh[N\alpha]-\cosh[p\alpha])-4{\sinh}^2\alpha ((-1)^N\cos[N\theta]-\cosh[N\alpha])\right.\nn \\
&+&\left. 4\sinh\alpha\sinh[N\alpha](A-2\cosh\alpha)\right)/(2{\sinh}^2\alpha)~~~{\rm with}~~ e^{\pm \alpha} =\frac{B}{2} \pm (\f{B^2}{4} -1)^{1/2}.\eea
with $p=N_2-N_1$. Similarly, the co-factor can be evaluated following the above trick. Here we find first $C_{1M}$ and calculate ${|C_{1M}|}^2$ which is relevant to determine conductance $G(\phi)$ of the asymmetric ring between the drain and source contacts.
\bea
{|C_{1M}|}^2&=&2\left[\cosh N\alpha_R \cosh p\alpha_R -\cos N\alpha_I \cos p\alpha_I + (-1)^N\{\cos N\theta~(\cos p\alpha_I \cosh N\alpha_R -\cos N\alpha_I \right. \nn \\
&& \left.\cosh p\alpha_R)+\sin N\theta~ (\sin p\alpha_I \sinh N\alpha_R -\sin N\alpha_I \sinh p\alpha_R )\}\right]/(\cosh 2\alpha_R -\cos 2\alpha_I ).~
\eea 
where $\alpha_R~{\rm and}~\alpha_I$ are respectively real and imaginary part of $\alpha$. For $\g'=0$, the real part of $\alpha$ vanishes and the coefficient of $\sin N\theta$ in ${|C_{1M}|}^2$ also disappears. We denote, ${{|C_{1M}|}^2}_{\g'=0}$ by ${|C^0_{1M}|}^2$ and $|{\Delta_N}|_{\g'=0}$ by $|{\Delta}^0_N|$.

Finally we evaluate the Green's function of Eq.~(\ref{mul}), where a single B{\"u}ttiker probe is coupled to a middle site (l) of the open ring. Here again $Z_{11}=Z^+_{MM}=A(\om)$, but all other diagonal terms are $\hbar \om/\g$ except $Z_{ll}=B(\om)$. The off-diagonal terms remain the same as before. Following the above method we calculate the Green's function ($l<M$)
\bea
G^+_{l1}&=&\f{(-1)^{l+1}\hbar}{2\g \Delta'_N \sinh^2\alpha'}\left[e^{i (l-1)\theta}\{B(\cosh[(N-l+1)\alpha']-\cosh[(r+1)\alpha'])- 2\cosh[(N-l)\alpha']\right. \nn \\
&& \left.+\cosh[r\alpha']+\cosh[(r+2)\alpha']\}+(-1)^{N}e^{i (l-N-1)\theta}(\cosh[l\alpha']-\cosh[(l-2)\alpha'])\right] \\
&& {\rm with}~~~e^{\pm \alpha'}=\f{\hbar\om}{2\g} \pm (\f{{\hbar}^2{\om}^2}{16\g^2}-1)^{1/2}~~,\nn 
\eea
where $r=N-2M+l$. In this case, we do not require to determine $\Delta'_N$, the determinant of $Z$, as it gets cancelled in  Eq.~(\ref{mul}). Similarly $G^+_{lM}$ can be evaluated. 


\end{document}